\documentclass[aps,prd,preprint,nofootinbib]{revtex4-1}

\newcommand{\beq}{\begin{equation}}                                             
\newcommand{\eeq}{\end{equation}}     
\newcommand{\bea}{\begin{eqnarray}}                                             
\newcommand{\eea}{\end{eqnarray}}     

\newcommand{\na}{\tilde{\chi}^0_1}
\newcommand{\nb}{\tilde{\chi}^0_2}
\newcommand{\ca}{\tilde{\chi}^\pm_1}
\newcommand{\cpa}{\tilde{\chi}^{+}_1}
\newcommand{\cma}{\tilde{\chi}^{-}_1}

\usepackage{graphicx}

\begin{document} 

\preprint{MS-TP-20-23}

\title{Higgsino and gaugino pair production at the LHC with aNNLO+NNLL precision}

\author{Juri Fiaschi}
\email{fiaschi@uni-muenster.de}
\author{Michael Klasen}
\email{michael.klasen@uni-muenster.de}

\affiliation{Institut f\"ur Theoretische Physik, Westf\"alische Wilhelms-Universit\"at M\"unster, Wilhelm-Klemm-Stra\ss{}e 9, 48149 M\"unster, Germany}


\begin{abstract}
  We present a calculation of higgsino and gaugino pair production at the LHC
  at next-to-next-to-leading logarithmic (NNLL) accuracy, matched to approximate
  next-to-next-to-leading order (aNNLO) QCD corrections. We briefly review the
  formalism for the resummation of large threshold logarithms and highlight the
  analytical results required at aNNLO+NNLO accuracy. Our numerical results are
  found to depend on the mass and nature
  of the produced charginos and neutralinos. The differential and total cross
  sections for light higgsinos, which like sleptons are produced mostly at small
  $x$ and in the $s$-channel, are found to be again moderately increased with
  respect to our previous results. The differential and total cross sections for
  gauginos are, however, not increased any more due to the fact that gauginos, like
  squarks, are now constrained by ATLAS and CMS to be heavier than about 1 TeV,
  so that also $t$- and $u$-channels play an important role. The valence quarks probed
  at large $x$ then also induce substantially different cross sections for positively
  and negatively charged gauginos. The higgsino and gaugino cross sections are
  both further stabilized at aNNLO+NNLL with respect to the variation of
  renormalization and factorization scales. We also now take mixing in the squark
  sector into account and study the dependence of the total cross sections on
  the squark and gluino masses as well as the trilinear coupling controlling the
  mixing in particular in the sbottom sector.
\end{abstract}

\maketitle

\section{Introduction}
\label{sec:1}

The Minimal Supersymmetric (SUSY) Standard Model (MSSM) is a theoretically and
phenomenologically well motivated extension of the Standard Model (SM) of particle
physics, that can solve a significant number of shortcomings of this model
\cite{Nilles:1983ge,Haber:1984rc}. Important examples in this respect are the
stabilization of the Higgs boson mass and the unification of strong and
electroweak forces at high scales. The MSSM predicts fermionic partners of the
neutral and charged gauge and Higgs bosons called gauginos and higgsinos, which
are typically among the lightest SUSY particles \cite{AguilarSaavedra:2005pw}.
The lightest neutral mass eigenstate, the lightest neutralino, is one of the best
studied dark matter candidates \cite{Ellis:1983ew,Ellis:1999mm,Klasen:2015uma,Herrmann:2009wk,%
Herrmann:2009mp,Harz:2012fz,Harz:2014tma,Harz:2014gaa,Branahl:2019yot}. Heavier
neutralinos and charginos decay typically into multilepton final states and
missing transverse momentum. Searches for higgsino- \cite{Sirunyan:2018iwl,%
Aad:2019qnd,Aaboud:2018zeb,CMS:2018qsc,Heister:2002mn,Abdallah:2003xe} or
gaugino-like particles \cite{Aaboud:2018jiw,Sirunyan:2018lul,Aad:2019vnb,%
Aad:2019vvi,Aad:2019vvf,Sirunyan:2018ubx} are therefore important physics goals at
the LHC. They are often carried out in the framework of simplified models
\cite{Alwall:2008ag,Calibbi:2014lga}. Care must, however, be taken that the
theoretical assumptions are not overly simplified \cite{Fuks:2017rio}.

Experimental measurements of supersymmetric (SUSY) production cross sections at
past and future runs of the LHC require precise theoretical calculations at the
level of next-to-leading order (NLO) QCD and beyond \cite{Beenakker:1996ch,%
Beenakker:1997ut,Beenakker:1999xh,Berger:1999mc,Berger:2000iu,Spira:2002rd,%
Jin:2003ez,Binoth:2011xi}. In the perturbative expansion, logarithmically
enhanced terms appear beyond leading order in the strong coupling constant
$\alpha_s$, whose contributions can be sizeable close to production threshold or
at small transverse momentum of the produced SUSY particle pair. Their effect on
neutralino, chargino \cite{Li:2007ih,Debove:2009ia,Debove:2010kf,Debove:2011xj,%
Fuks:2012qx,Fuks:2016vdc}, slepton \cite{Yang:2005ts,Broggio:2011bd,%
Bozzi:2006fw,Bozzi:2007qr,Bozzi:2007tea,Fuks:2013lya}, squark, gluino
\cite{Beenakker:2014sma,Borschensky:2014cia,Beneke:2016kvz,Beenakker:2016lwe},
stop \cite{Broggio:2013cia,Beenakker:2016gmf} and also new gauge boson production
\cite{Fuks:2007gk,Jezo:2014wra,Mitra:2016kov,Klasen:2016qux} has been taken into
account to all orders with resummation techniques to next-to-leading logarithmic
(NLL) accuracy and beyond. The results for the electroweak production channels
have been made publicly available with the code RESUMMINO \cite{Fuks:2013vua} and
are regularly employed in the experimental analyses by ATLAS \cite{Aad:2019vvf}
and CMS \cite{Sirunyan:2018iwl}. Predictions have also recently been made for the
high-luminosity (HL) and high-energy (HE) phases of the LHC
\cite{CidVidal:2018eel}. The effect of higher order QCD corrections is generally
to enhance the theoretical estimations for the cross sections, while on the other
hand they reduce the dependence of the results on the choice of the unphysical
renormalization and factorization scales. Together with resummation-improved
parton density functions (PDFs) \cite{Bonvini:2015ira}, also the PDF uncertainty
can in principle be reduced \cite{Beenakker:2015rna,Fiaschi:2018xdm,%
Fiaschi:2018buk,Fiaschi:2018hgm,Fiaschi:2019etm}, even though in practice these
PDFs must currently be fitted to smaller data sets than global NLO analyses and
thus still have larger errors.

In this paper, we take our precision calculations for higgsino and gaugino pair
production to the next level by resumming not only the leading and
next-to-leading logarithms (NLL), but also the next-to-next-to-leading logarithms
(NNLL) and matching them not only to the full NLO QCD and SUSY-QCD corrections,
but also an approximate next-to-next-to-leading order (aNNLO) calculation in QCD.
The corresponding analytical formulae are available in the literature
\cite{Vogt:2000ci,Kidonakis:2003tx, Kidonakis:2007ww,Fiaschi:2019zgh}, so that we
collect here only the most important results required at NNLL accuracy. Similar
calculations, based on full NLO SUSY-QCD and aNNLO QCD calculations
\cite{Beenakker:1996ch,Beenakker:1997ut}, have also been performed previously for
sleptons \cite{Fiaschi:2019zgh} as well as for squarks, gluinos
\cite{Beenakker:2014sma} and stops \cite{Beenakker:2016gmf} and are available
through the public codes RESUMMINO \cite{Fuks:2013vua} and NNLL-fast
\cite{Beenakker:2016lwe}. Other groups have employed soft-collinear effective
theory for sleptons \cite{Broggio:2011bd}, squarks \cite{Beneke:2010da}, gluinos
\cite{Beneke:2016kvz} and stops \cite{Broggio:2013uba,Broggio:2013cia} with
similar conclusions.

The paper is organized as follows: In Sec.\ \ref{sec:2}, we present our analytical
approach and in particular how threshold logarithms can be resummed at NNLL
accuracy, matched to a fixed-order calculation up to NNLO and how the PDFs and
hadronic cross sections are transformed to and from Mellin space. Our numerical
results for the production of relatively light higgsino pairs are contained in
Sec.\ \ref{sec:3}. This section starts with a discussion of the QCD and SUSY input
parameters, followed by a demonstration of how the NNLL and aNNLO contributions
affect the differential cross section at small and large invariant masses. We
then show the effects of the new contributions on the total cross section and its
dependence on the factorization and renormalization scales. We also discuss the
dependence on other SUSY parameters like the squark and gluino masses and the
trilinear coupling governing squark mixing in the bottom sector. Numerical results
for the pair production of heavier gauginos are described in a similar way in
Sec.\ \ref{sec:4}. The ensuing conclusions are presented in Sec.\ \ref{sec:5}.

\section{Analytical approach}
\label{sec:2}

The hadronic invariant mass distribution for the pair production of neutralinos
and charginos
\bea
 M^2{d\sigma_{AB}\over dM^2}(\tau)&=&\sum_{a,b}\int_0^1dx_adx_bdz
 \left[x_af_{a/A}(x_a,\mu_F^2)\right]
 \left[x_bf_{b/B}(x_b,\mu_F^2)\right]\nonumber\\
 &&\times\left[z\sigma_{ab}(z,M^2,\mu_R^2,\mu_F^2)\right]\delta(\tau-x_ax_bz)
 \label{eq:2.1}
\eea
requires the convolution of parton density functions (PDFs) $f_{a,b/A,B}$ with the
partonic cross section $\sigma_{ab}$. The former depend on $x_{a,b}$, the
longitudinal momentum fractions of the partons $a$ and $b$ in the colliding
hadrons $A$ and $B$, and the factorization scale $\mu_F$. The latter is a
function of the squared invariant mass of the produced neutralinos or charginos
$M^2$, its ratio $z=M^2/s$ to the partonic center-of-mass energy $s$, and the
renormalization and factorization scales $\mu_R$ and $\mu_F$. In contrast to the
leading order (LO) cross section \cite{Dawson:1983fw,Bozzi:2007me} and the virtual
next-to-leading order (NLO) corrections, which are proportional to $\delta(1-z)$
\cite{Beenakker:1999xh}, the kinematic mismatch in the cancellation of infrared
divergences among the virtual and real corrections of order $n$ introduces large
logarithmic remainders proportional to
\bea
 \alpha_s^n(\mu_R^2) \left[{\ln^m(1-z)\over1-z}\right]_+ &,& {\rm where} \
 m\leq2n-1.
 \label{eq:2.2}
\eea
Close to threshold ($z\to1$), they spoil the convergence of the perturbative
series in $\alpha_s$ and therefore have to be resummed to all orders
\cite{Sterman:1986aj,Catani:1989ne}.

After performing a Mellin transformation of the PDFs and partonic cross section in
Eq.\ (\ref{eq:2.1}), the hadronic cross section $\sigma_{AB}$ factorizes, the
singular terms in Eq.\ (\ref{eq:2.2}) turn into large logarithms of the Mellin
variable $N$,
\bea
 \left[{\ln^m(1-z)\over1-z}\right]_+ & \to & \ln^{m+1}N+\dots,
\eea
and the partonic cross section $\sigma_{ab}$ can be written in the exponentiated
form
\beq
 \sigma_{ab}^{\rm (res.)}(N,M^2,\mu_R^2,\mu_F^2) =
 H_{ab}(M^2,\mu_R^2,\mu_F^2)\exp[G_{ab}(N,M^2,\mu_R^2,\mu_F^2)]
 + \mathcal{O}\left(\frac{1}{N}\right).
 \label{eq:2.5}
\eeq
Here, the exponent $G_{ab}$ is universal and contains all the logarithmically
enhanced contributions in the Mellin variable $N$, while the hard function
$H_{ab}$ is independent of $N$, though process-dependent.

Up to next-to-next-to-leading logarithmic (NNLL) accuracy, the exponent $G_{ab}$
can be written as
\beq
 G_{ab}(N,M^2,\mu_R^2,\mu_F^2) =
 L G_{ab}^{(1)}(\lambda) +
 G_{ab}^{(2)}(\lambda, M^2, \mu_R^2,\mu_F^2) +
 \alpha_s G_{ab}^{(3)}(\lambda, M^2, \mu_R^2, \mu_F^2),
\eeq
where $\lambda = \alpha_s b_0 L$ and $L = \ln\bar{N} = \ln(Ne^{\gamma_E})$. For
Drell-Yan-like processes such as slepton or higgsino and gaugino pair production
initiated by quarks and antiquarks only, the coefficients $G^{(i)}_{ab}=g_a^{(i)}+
g_b^{(i)}$ with $a=b=q$ can be found up to next-to-leading logarithmic (NLL)
accuracy in Refs.~\cite{Bozzi:2007qr, Debove:2010kf}. In addition to the LL and NLL
terms $g_q^{(1)}$ and $g_q^{(2)}$, one needs at NNLL also \cite{Vogt:2000ci}
\bea
 g_q^{(3)}(\lambda) &=& \frac{A^{(1)} b_1^2}{2 \pi b_0^4} \frac{1}{1 - 2 \lambda} \left[2 \lambda^2 + 2 \lambda \ln(1-2\lambda) + \frac{1}{2}\ln^2(1-2\lambda)\right] \nonumber\\
&+& \frac{A^{(1)} b_2}{2 \pi b_0^3} \left[2 \lambda + \ln(1-2\lambda) + \frac{2\lambda^2}{1 - 2 \lambda}\right] +\frac{2A^{(1)}}{\pi} \zeta_2 \frac{\lambda}{1 - 2 \lambda} \nonumber\\
&-&  \frac{A^{(2)} b_1}{(2 \pi)^2 b_0^3} \frac{1}{1 - 2 \lambda} \left[2 \lambda^2 + 2 \lambda + \ln(1-2\lambda)\right] +\frac{A^{(3)}}{\pi^3 b_0^2} \frac{\lambda^2}{1 - 2 \lambda} - \frac{D^{(2)}}{2 \pi^2 b_0} \frac{\lambda}{1 - 2 \lambda}  \nonumber\\
&+&  \frac{A^{(1)} b_1}{2 \pi b_0^2} \frac{1}{1 - 2 \lambda} \left[2 \lambda + \ln(1-2\lambda)\right]\ln\left(\frac{M^2}{\mu_R^2}\right) +\frac{A^{(1)}}{2 \pi} \left[\frac{\lambda}{1 - 2 \lambda}\ln^2\left(\frac{M^2}{\mu_R^2}\right) - \lambda\ln^2\left(\frac{\mu_F^2}{\mu_R^2}\right)\right]\nonumber\\
&-& \frac{A^{(2)}}{2 \pi^2 b_0} \left[\frac{\lambda}{1 - 2 \lambda}\ln\left(\frac{M^2}{\mu_R^2}\right) - \lambda\ln\left(\frac{\mu_F^2}{\mu_R^2}\right)\right].
\eea
Here, the new coefficients required at NNLL are given by \cite{Moch:2005ba}
\bea
 A^{(3)} &=& \frac{1}{2} C_F \Big[C_A^2 \left(\frac{245}{24} - \frac{67}{9}\zeta_2 + \frac{11}{6}\zeta_3 + \frac{11}{5}\zeta_2^2 \right) +C_F n_f \left(2\zeta_3 - \frac{55}{24} \right) \nonumber \\
 &+& C_A n_f \left(\frac{10}{9}\zeta_2 - \frac{7}{3} \zeta_3 - \frac{209}{108} \right) - \frac{n_f^2}{27} \Big]
\eea
and \cite{Vogt:2000ci} 
\beq
 D^{(2)}=2C_F\left[C_A\left( -{101\over27}+{11\over3}\zeta_2+{7\over2}\zeta_3\right)
 +n_f\left({14\over27}-{2\over3}\zeta_2\right)\right].
\eeq
The coefficients of the QCD $\beta$-function are denoted by $b_n=\beta_n/(2\pi)^{n+1}$
\cite{Tarasov:1980au,Larin:1993tp}, the QCD color factors are $C_A=N_C=3$ and
$C_F=(N^2-1)/(2N_C)=4/3$, and the number of active quark flavors is $n_f=5$.

The hard $N$-independent part of the Mellin-transformed partonic cross section
in Eq.\ (\ref{eq:2.5})
\beq
 H_{ab} (M^2,\mu_R^2,\mu_F^2) = \sigma^{(0)}_{ab} \mathcal{C}_{ab}(M^2,\mu_R^2,\mu_F^2)
\eeq
can be perturbatively expanded in terms of the Mellin-transformed LO cross section
$\sigma^{(0)}_{ab}$ and
\beq
 \mathcal{C}_{ab}(M^2,\mu_R^2,\mu_F^2) =
 \sum_{n=0}\left(\frac{\alpha_s}{2\pi}\right)^n \mathcal{C}_{ab}^{(n)}(M^2,\mu_R^2,\mu_F^2),
\eeq
where the hard matching coefficients
\beq
 \mathcal{C}_{ab}^{(n)}(M^2,\mu_R^2,\mu_F^2) =
 \left(\frac{2\pi}{\alpha_s}\right)^n \left[\frac{\sigma^{(n)}_{ab}}{\sigma^{(0)}_{ab}}\right]_{\rm N-ind.}
 \label{eq:2.17}
\eeq
are obtained from the finite ($N$-independent) terms in the ratio of the $n$-th order
cross section over the LO one. The QCD part of the coefficient required at
next-to-next-to-leading order (NNLO) in pair-invariant mass kinematics is given by
\cite{Kidonakis:2003tx, Kidonakis:2007ww}
\bea
 \mathcal{C}_{q\bar{q}}^{(2)} &=& \frac{C_F}{720} \bigg\{ 5(-4605 C_A + 4599 C_F + 762 n_f) + 20 \pi^2 (188 C_A - 297 C_F - 32 n_f) \\ \nonumber
 &-& 92 \pi^4 (C_A - 6 C_F) + 180 (11 C_A + 18 C_F - 2 n_f)\log^2\left(\frac{\mu_F^2}{M^2}\right) \\ \nonumber
 &-& 160 (11 C_A - 2 n_f)(6 - \pi^2)\log\left(\frac{\mu_R^2}{M^2}\right) + 80 (151 C_A - 135 C_F + 2 n_f) \zeta_3 \\ \nonumber
 &+& 20 \log\left(\frac{\mu_F^2}{M^2}\right) \bigg[-51 C_A + 837 C_F + 6 n_f - 4 \pi^2 (11 C_A + 27 C_F - 2 n_f) \\ \nonumber
 &+& (-198 C_A + 36 n_f)\log\left(\frac{\mu_R^2}{M^2}\right) + 216 (C_A - 2 C_F)\zeta_3\bigg]  \bigg\}.
\eea
It allows to improve the resummation of logarithmically enhanced contributions, since
also beyond NNLO in $\alpha_s$ the finite terms are multiplied by threshold logarithms.

While in the limit of large $N$ the cross section is clearly dominated by terms of
$\mathcal{O}(\ln^{2} N)$, $\mathcal{O}(\ln N)$ and $\mathcal{O}(1)$, some of the terms
suppressed by powers of $1/N$ are multiplied by powers of $\ln N$ and can thus also
have a non-negligible effect \cite{Kramer:1996iq,Catani:2001ic,Catani:2003zt,%
Beneke:2018gvs,Beneke:2019oqx,Beneke:2019mua,Bahjat-Abbas:2019fqa}. This collinear
improvement is taken into account in RESUMMINO for higgsino, gaugino and
slepton pair production \cite{Debove:2010kf,Bozzi:2007qr}. Since we have given a
detailed discussion of the collinear improvement as well as the theoretical status
of exponentiating contributions in the $\mathcal{C}$-function \cite{Eynck:2003fn,%
Duhr:2020seh,H.:2020ecd} in a previous publication \cite{Fiaschi:2019zgh}, we do
not repeat it here. The same holds for the matching of resummed cross section
$\sigma_{ab}^{\rm (res.)}$, valid near threshold, and the normal perturbative
calculation $\sigma_{ab}^{\rm (f.o.)}$, valid outside this region, by adding both
results and subtracting the overlap $\sigma_{ab}^{\rm (exp.)}$, i.e.\ the resummed
cross section re-expanded to NNLO. We therefore give here only the
contributions to the new coefficient
\bea
 \mathcal{K}^{(2)} &=& \mathcal{K}^{(2,1)} L + \mathcal{K}^{(2,2)} L^2 + \mathcal{K}^{(2,3)} L^3 + \mathcal{K}^{(2,4)} L^4
\eea
required at NNLO in the expanded cross section
\bea
 \sigma^{\rm (exp.)}_{ab}(N,M^2\!\!&\!,\!&\!\mu_R^2,\mu_F^2) =
 \sigma^{(0)}_{ab} \mathcal{C}_{ab}(M^2,\mu_R^2,\mu_F^2) \exp[G_{ab}(N,M^2,\mu_R^2,\mu_F^2)] \nonumber \\
 &=& \sigma^{(0)}_{ab} \left[1 + \left(\frac{\alpha_s}{2\pi}\right) \mathcal{C}_{ab}^{(1)} + \left(\frac{\alpha_s}{2\pi}\right)^2 \mathcal{C}_{ab}^{(2)} +\dots\right]\left[1 + \left(\frac{\alpha_s}{2\pi}\right) \mathcal{K}^{(1)} + \left(\frac{\alpha_s}{2\pi}\right)^2 \mathcal{K}^{(2)} +\dots\right] \nonumber \\
 &=& \sigma^{(0)}_{ab}\left[1 + \left(\frac{\alpha_s}{2\pi}\right) \left(\mathcal{C}_{ab}^{(1)} + \mathcal{K}^{(1)}\right) + \left(\frac{\alpha_s}{2\pi}\right)^2 \left(\mathcal{C}_{ab}^{(2)} + \mathcal{K}^{(2)} + \mathcal{C}_{ab}^{(1)}\mathcal{K}^{(1)}\right) + \dots\right],
\eea
which read explicitly\cite{Kidonakis:2003tx, Kidonakis:2007ww}
\bea
 \mathcal{K}^{(2,1)} &=& -\frac{C_F}{27} \bigg\{56 n_f - 404 C_A + 3\log\left(\frac{\mu_F^2}{s}\right) \bigg[20n_f + 2 C_A (-67 + 3\pi^2) \nonumber\\
 &+& 3 (11C_A - 2 n_f) \left(\log\left(\frac{\mu_F^2}{\mu_R^2}\right) - \log\left(\frac{\mu_R^2}{s}\right)\right)\bigg] + 378 C_A \zeta_3 \bigg\}, \\
 \mathcal{K}^{(2,2)} &=& \frac{2}{9} C_F \bigg[-10 n_f + 67 C_A - 3 C_A \pi^2 + 36 C_F \log^2\left(\frac{\mu_F^2}{s}\right) \nonumber \\
 &+& (33 C_A - 6 n_f) \log\left(\frac{\mu_R^2}{s}\right) \bigg], \\
 \mathcal{K}^{(2,3)} &=& \frac{4}{9} C_F \left[11 C_A - 2 n_f + 36 C_F \log\left(\frac{\mu_F^2}{s}\right)\right], \\
 \mathcal{K}^{(2,4)} &=& 8 C_F^2.
\eea
The SUSY-QCD (squark-gluino loop) corrections are only matched at NLO, since they
are not known beyond this order \cite{Beenakker:1999xh}. In this sense, our results
are accurate to approximate NNLO (aNNLO) plus NNLL precision. This approximation is
justified by the fact that the SUSY-QCD corrections are subdominant due to the
large squark and gluino masses. A detailed description of the inverse Mellin
transform
\bea
 \label{eq:2.51}
 M^2{d\sigma_{AB}\over d M^2}(\tau)&=&{1\over2\pi i}\int_{{\cal C}_N} d N 
g \tau^{-N} M^2{d\sigma_{AB}(N)\over d M^2},
\eea
that has to be performed for the resummed and the perturbatively expanded results
in Mellin space can be found in Ref.\ \cite{Fiaschi:2019zgh}.

\section{Numerical results for higgsino pair production}
\label{sec:3}

Naturalness arguments on the spectrum of SUSY theories require the masses of
higgsinos to be small, i.e. below the TeV scale, and the lightest SUSY particle
(LSP) $\tilde{\chi}^0_1$, the lightest chargino ($\tilde{\chi}^\pm_1$) and the
next-to-lightest neutralino ($\tilde{\chi}^0_2)$ to be close in mass. Experimental
analyses with the largest sensitivity to this kind of compressed scenario consider
three main processes, which all lead to signatures with soft leptons and moderate
missing transverse momentum in the final state \cite{Sirunyan:2018iwl}. The first
two processes are the associated production of a positively or negatively charged
$\tilde{\chi}^\pm_1$ and a $\tilde{\chi}^0_2$, while in the third process
a pair of charginos ($\tilde{\chi}^+_1\tilde{\chi}^-_1$) is produced. The heavier
neutralino $\tilde{\chi}^0_2$ and the charginos $\tilde{\chi}^\pm_1$ will decay to
the lighter $\tilde{\chi}^0_1$ through an off-shell $Z$ or $W^\pm$ boson,
respectively. Since the decay products are expected to be soft because of the
compressed spectrum, a jet with large transverse momentum produced through initial
state radiation (ISR) can enhance the discriminating power with respect to SM
processes \cite{Sirunyan:2018iwl}.

Based on an integrated LHC luminosity of 139 (36) fb$^{-1}$, the ATLAS (CMS)
collaboration have excluded pure, mass-degenerate higgsino pairs $\ca\nb$ up to
193 (168) GeV, when they decay to 9 (20) GeV lighter $\na$'s and electroweak $W$
and $Z$ gauge bosons \cite{Aad:2019qnd,Sirunyan:2018iwl}. For general
gauge-mediated SUSY breaking models, the limits set by the ATLAS collaboration are
somewhat stronger with 295 GeV for mass-degenerate higgsinos including the $\na$
that decay to $Z$ (or $h$) bosons and almost massless gravitinos $\tilde{G}$
\cite{Aaboud:2018zeb}. In the high-luminosity phase of the LHC (HL-LHC) with 3000
fb$^{-1}$ at a center-of-mass energy 14 TeV, the mass reach is expected to extend
to 360 GeV \cite{CMS:2018qsc}. For the invariant-mass distributions we therefore
adopt a default $\nb$ ($\ca$) mass of 208 (203) GeV, while for the total cross
section analysis we vary the $\ca$ mass between the LEP limit of 103.5 GeV, valid
for a mass splitting with the $\na$ of at least 3 GeV \cite{Heister:2002mn,%
  Abdallah:2003xe}, and 500 GeV.

Gluinos enter only at NLO in virtual loop diagrams, so that their masses play a
subdominant role. Squarks appear already at LO in the $t$- and $u$-channel propagators,
but since light higgsinos are mostly produced in the $s$-channel, their masses also
have little influence, as does the trilinear coupling $A_0$ determining mixing in
the sbottom sector. We adopt a squark and gluino mass of 1.3 TeV as our default
value, which is still allowed for not too large mass differences with the lightest
neutralino, even though the most stringent ATLAS (CMS) mass limits already reach
1.94 (1.63) and 2.35 (2.31) TeV \cite{ATLAS:2019vcq,Sirunyan:2019ctn}.

In the following, we compute the cross sections for the aforementioned processes
at LO, NLO, NLO+NLL and aNNLO+NNLL adopting CT14 PDFs at LO, NLO and NNLO for
consistency \cite{Dulat:2015mca}. The spectra with the specific characteristics
of MSSM scenarios have been obtained with the public code SPheno 4.0.3
\cite{Porod:2003um,Porod:2011nf}, following the considerations in
Ref.~\cite{Fuks:2017rio}. In particular, light higgsino-like neutralinos and
charginos $\tilde{\chi}^0_1$, $\tilde{\chi}^\pm_1$ and $\tilde{\chi}^0_2$ of
masses similar to the higgsino mass parameter $\mu$ can be obtained by setting
this parameter to $\mu\leq M_1 = M_2$, i.e.\ below the bino and wino mass
parameters $M_1$ and $M_2$. We set $\tan\beta=30$ and choose 
$\mu$ between 100 GeV and 500 GeV in
order to stay (not too far) above the experimental exclusion limits, while
our choice $M_{1,2}$ = 1 TeV ensures a large higgsino content and mass
splittings of the order of 5 GeV (i.e. $m_{\tilde{\chi}^0_2}-
m_{\tilde{\chi}^\pm_1} \approx m_{\tilde{\chi}^\pm_1} - m_{\tilde{\chi}^0_1}
\approx$ 5 GeV). Our calculations of differential and total cross sections are
performed using RESUMMINO \cite{Fuks:2013vua} interfaced with LHAPDF6
\cite{Buckley:2014ana} for the interpolation of the PDF grids. The SM parameters
have been chosen according to their current PDG values \cite{Tanabashi:2018oca},
and $\alpha_s(\mu_R)$ is computed in accordance with the corresponding CT14 PDF
fit.

\subsection{Invariant-mass distributions}

\begin{figure}
\begin{center}
\includegraphics[width=0.75\textwidth]{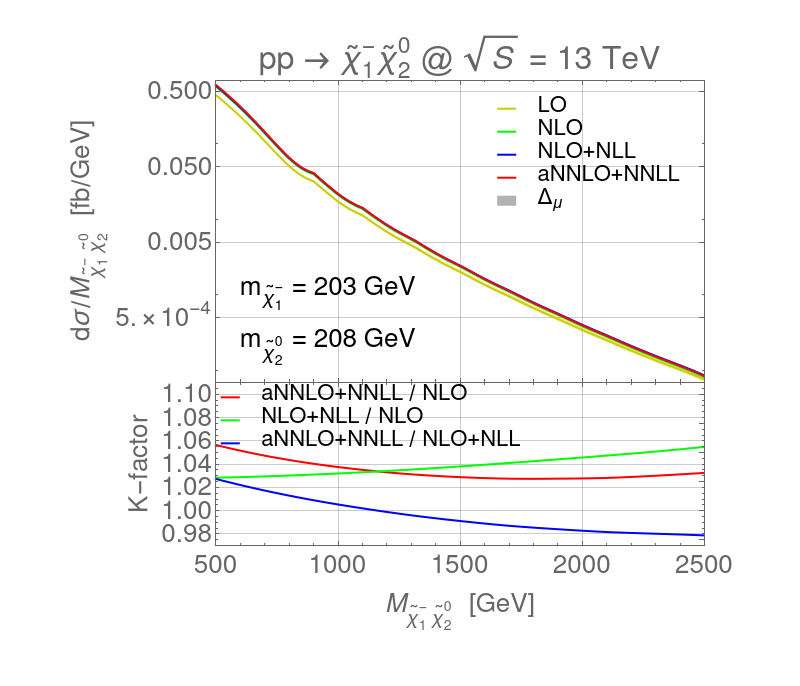}
\caption{Top: Invariant-mass distribution for the associated production of 
  charginos and neutralinos with masses of 203 GeV and 208 GeV
  at the LHC with a center-of-mass energy of $\sqrt{S}=13$~TeV. Shown are results
  at LO (yellow), NLO (green), NLO+NLL (blue) and aNNLO+NNLL (red) together with
  the corresponding scale uncertainties (shaded bands). Bottom: Ratios ($K$
  factors) of aNNLO+NNLL over NLO (red), NLO+NLL over NLO (green) and aNNLO+NNLL
  over NLO+NLL (blue) differential cross sections as a function of the invariant
  mass of the higgsino pair.}
\label{fig:1}
\end{center}
\end{figure}

We begin with the invariant-mass distribution for the associated production of a
higgsino-like lightest chargino and a higgsino-like second-lightest neutralino.
These differential cross sections
at LO (yellow), NLO (green), NLO+NLL (blue) and aNNLO+NLL (red curve) are shown
in the upper panel of Fig.~\ref{fig:1}. On a logarithmic scale, the uncertainties
(shaded bands) coming from variations of the renormalization and factorization
scales with the seven-point method (i.e.\ by relative factors of two, but not
four) about their central value, the average produced SUSY particle mass, are
barely visible, and we will study them in more detail in Fig.\ \ref{fig:2}
below. Also the relative impact of the higher order corrections is only clearly
visible with respect to the LO prediction, so that we have included a lower panel
showing the ratios ($K$ factors) of aNNLO+NLL over NLO (red), NLO+NLL over NLO
(green) and also aNNLO+NNLL over NLO+NLL (blue curve) differential cross sections.
While the NLO corrections have long been known to enhance the LO cross section by
about 30\% \cite{Beenakker:1999xh}, the NLL and NNLL corrections increase the NLO
cross section by another 3-5\% and $\pm2$ \%, respectively, showing a good
convergence of the perturbative series.

This is also demonstrated by the width of the combined scale uncertainty, shown
in Fig.\ \ref{fig:2} as shaded bands at NLO+NLL (blue) and aNNLO+NNLL (red).
\begin{figure}
\begin{center}
\includegraphics[width=0.75\textwidth]{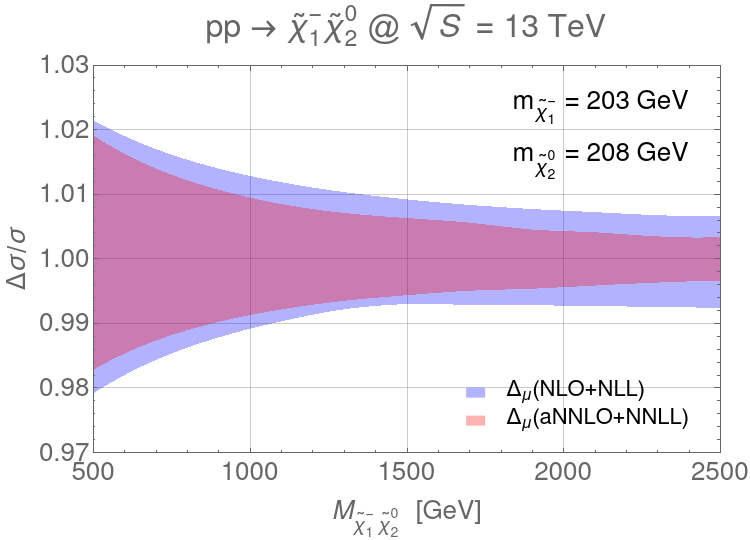}
\caption{Scale uncertainty of the invariant-mass distribution in Fig.\ \ref{fig:1}.
  Shown are the results at NLO+NLL (blue) and aNNLO+NNLL (red shaded band).}
\label{fig:2}
\end{center}
\end{figure}
At small and large invariant masses, this uncertainty shrinks from $\pm2.1$ to
1.8\% and from $\pm0.6$ to 0.4\%. As expected, resummation of large threshold
logarithms stabilizes the cross section more for large invariant masses, in
particular when the final state is mostly produced in the $s$-channel as it is
the case for light higgsinos and sleptons \cite{Fiaschi:2019zgh}.

\subsection{Total cross sections}

\begin{figure}
\begin{center}
\includegraphics[width=0.75\textwidth]{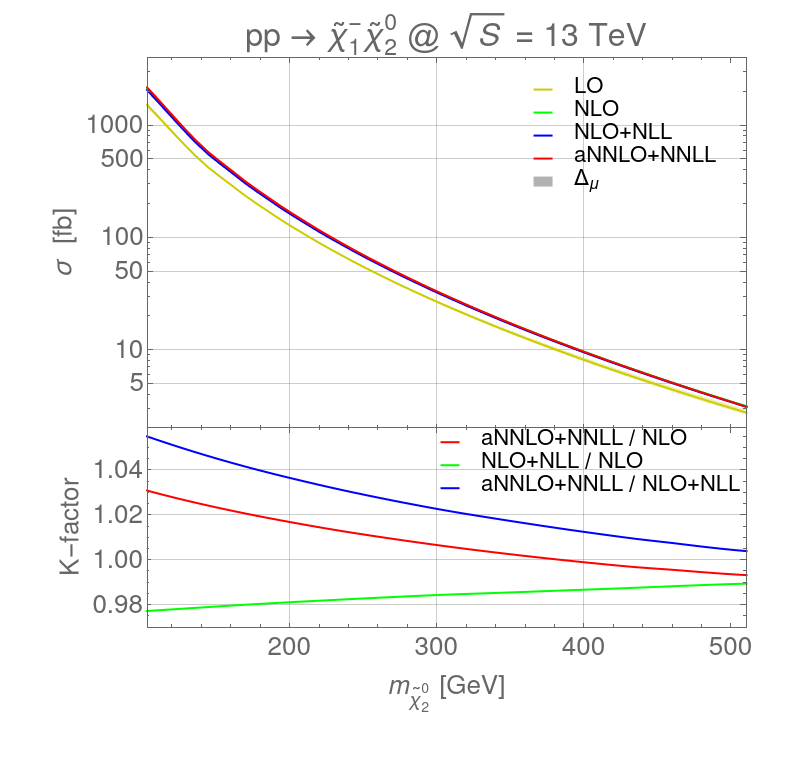}
\caption{Top: Total cross section for higgsino-like charginos and neutralinos at
  the LHC with a center-of-mass energy
  of $\sqrt{S}=13$ TeV as a function of the $\nb$ mass. Shown are results at LO
  (yellow), NLO (green), NLO+NLL (blue) and aNNLO+NNLL (red) together with the
  corresponding scale uncertainties (shaded bands). Bottom: Ratios ($K$
  factors) of aNNLO+NNLL over NLO (red), NLO+NLL over NLO (green) and aNNLO+NNLL
  over NLO+NLL (blue) total cross sections.}
\label{fig:3}
\end{center}
\end{figure}

The total cross section for the associated production of higgsino-like charginos
and neutralinos is shown in the upper panel of Fig.\ \ref{fig:3} at
LO (yellow), NLO (green), NLO+NLL (blue) and aNNLO+NLL (red curve). Again, only
the increase from LO to higher order is clearly visible on the logarithmic scale.
The enhancements from NLO to aNNLO+NNLL (red) and to NLO+NLL (green) as well as
their ratio (blue) are therefore shown in the lower panel. The aNNLO+NNLL corrections
increase the total cross section by up to 5\% for low higgsino masses, and the
perturbation series converges nicely for large higgsino masses.

\begin{figure}
\begin{center}
\includegraphics[width=0.49\textwidth]{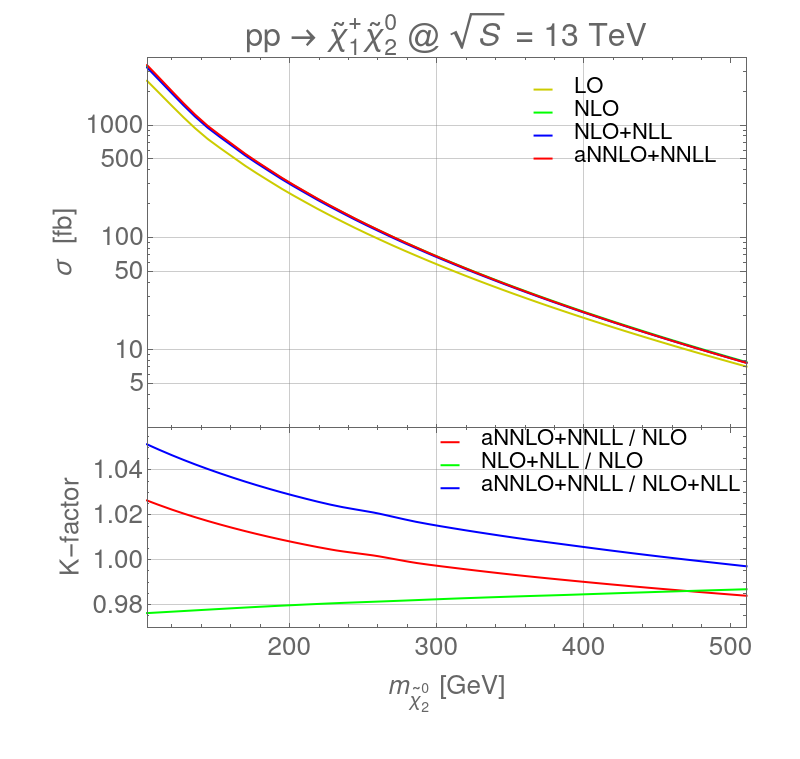}
\includegraphics[width=0.49\textwidth]{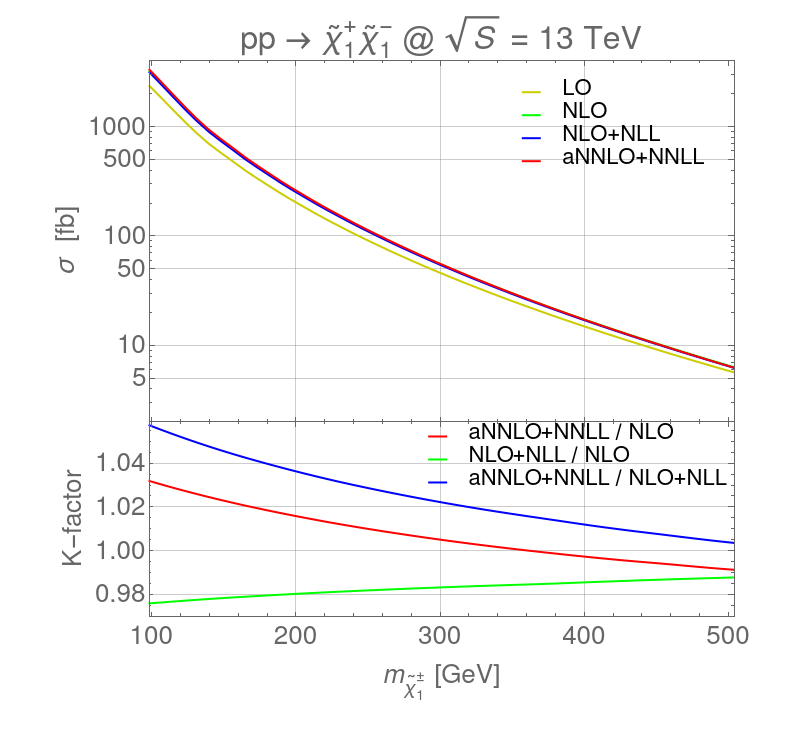}
\caption{Same as Fig.\ \ref{fig:3}, but for the associated production of a positively
  charged higgsino with the second-lightest neutralino (left) and for the pair
  production of charginos (right).}
\label{fig:4}
\end{center}
\end{figure}

The situation is very similar for the production of higgsino-like $\cpa\nb$ and
chargino pairs $\cpa\cma$, shown in Fig.\ \ref{fig:4}. 
The main difference is the absolute size of the total cross section, which at
a $pp$ collider like the LHC is largest for positively charged final states,
followed by neutral and negatively charged final states. The magnitude of the
difference depends on the $x$-range probed in the PDFs and thus on the
higgsino masses. It increases towards larger masses, where valence quarks play
a more important role.

\begin{figure}
\begin{center}
\includegraphics[width=0.49\textwidth]{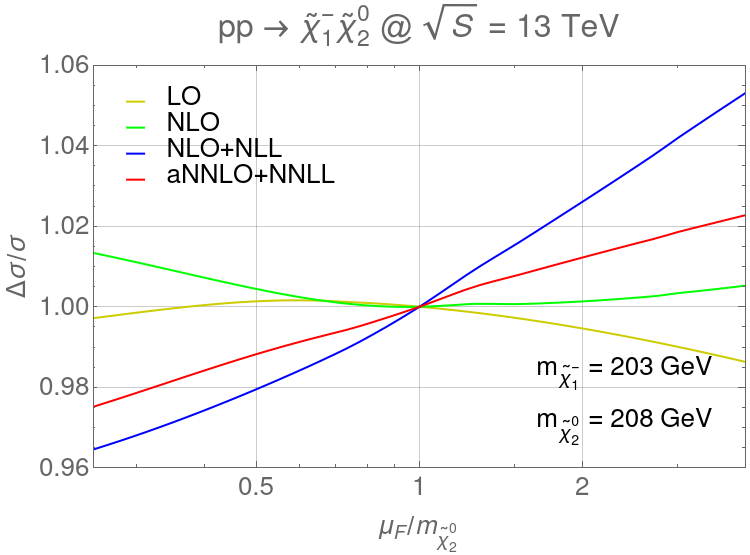}
\includegraphics[width=0.49\textwidth]{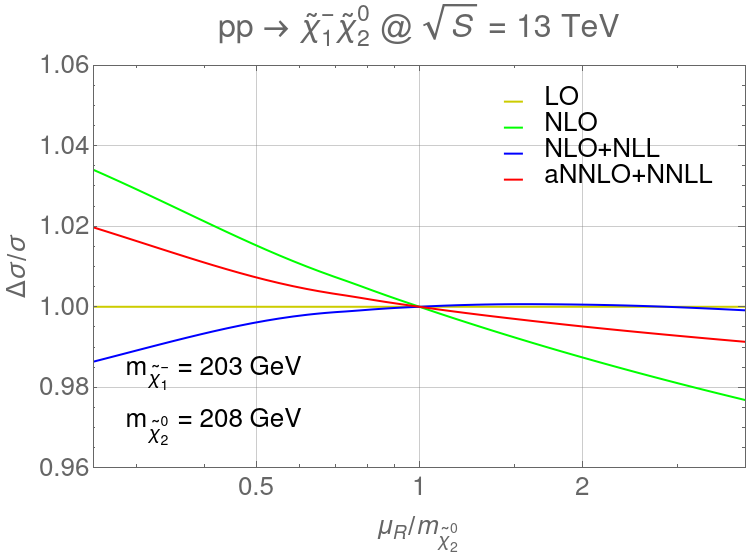}
\caption{Relative variation of the total cross section for higgsino-like
  chargino-neutralino pairs as a function of the factorization (top) and renormalization
  scale (bottom). Shown are results at LO (yellow), NLO (green), NLO+NLL
  (blue) and aNNLO+NNLL (red).}
\label{fig:5}
\end{center}
\end{figure}

\begin{figure}
\begin{center}
\includegraphics[width=0.75\textwidth]{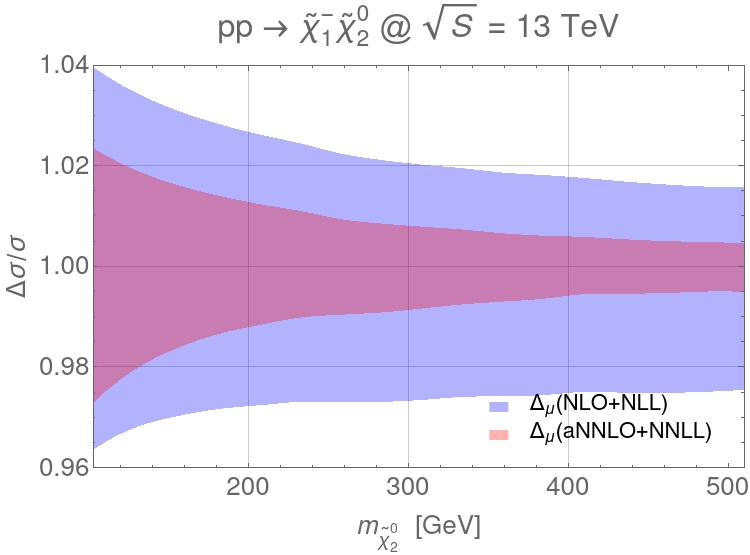}
\caption{Scale uncertainty of the total cross section for higgsino-like
  chargino-neutralino pairs at the LHC with a center-of-mass energy of
  $\sqrt{S}=13$ TeV as a function of the neutralino mass. Shown are the
  results at NLO+NLL (blue) and aNNLO+NNLL (red shaded band).}
\label{fig:6}
\end{center}
\end{figure}

The dependence of the total higgsino cross section on the factorization (top)
and renormalization (bottom) scales is studied individually in Fig.\ \ref{fig:5}.
While the latter is only introduced only at NLO (green), the former includes a
weak dependence from the PDFs already at LO (yellow). From NLO+NLL (blue) to
aNNLO+NNLL (red) one observes a reduction in particular for the factorization
scale. At these relatively low higgsino masses of 203 and 208 GeV, respectively,
both uncertainties still amount to about $\pm2$\%, while at NLO and even NLO+NLL
they could still reach about $\pm4$\%.
This is also reflected in Fig.\ \ref{fig:6}, where both uncertainties are varied
with the seven-point method and shown as a function of the higgsino mass. As
expected, the combined uncertainty is reduced for heavier higgsinos to a level
of about $\pm2$\% at NLO+NLL and only $\pm0.5$\% at aNNLO+NNLL. The situation
for higgsinos, which are mostly in the $s$-channel, is thus similar to the one
for sleptons \cite{Fiaschi:2019zgh}.

\begin{figure}
\begin{center}
\includegraphics[width=0.70\textwidth]{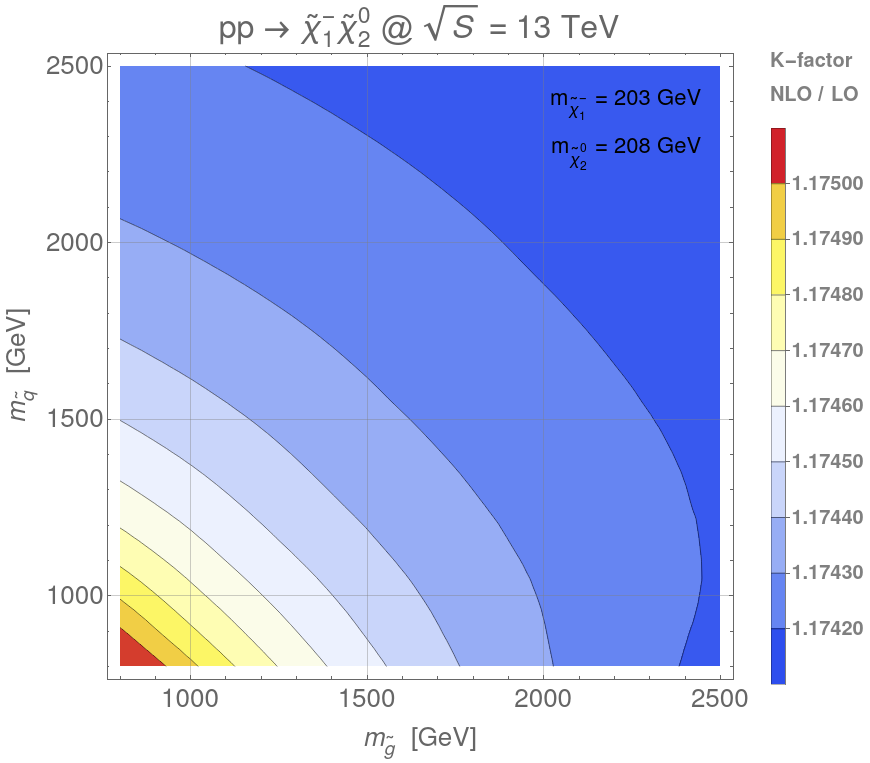}
\caption{Ratio ($K$ factor) of NLO over LO total cross sections (both
  with NLO PDFs) for higgsino pair production
  at the LHC with a center-of-mass energy of $\sqrt{S}=13$ TeV as a
  function of the squark and gluino masses.}
\label{fig:7}
\end{center}
\end{figure}

Since the $t$- and $u$-channels play a subdominant role for higgsinos, not only
the dependence on the gluino mass, introduced only at NLO, but also the one on
the squark masses should be weak. This can clearly be seen in Fig.\ \ref{fig:7},
where the ratio of the NLO (and similarly any other higher-order) cross section
over the LO one is shown in the squark-gluino mass plane. Overall, it varies by
less than one per mill. The gradient is along the diagonal and slightly steeper
when the squark and gluino masses are still relatively close to those of the
higgsinos.

\begin{figure}
\begin{center}
\includegraphics[width=0.62\textwidth]{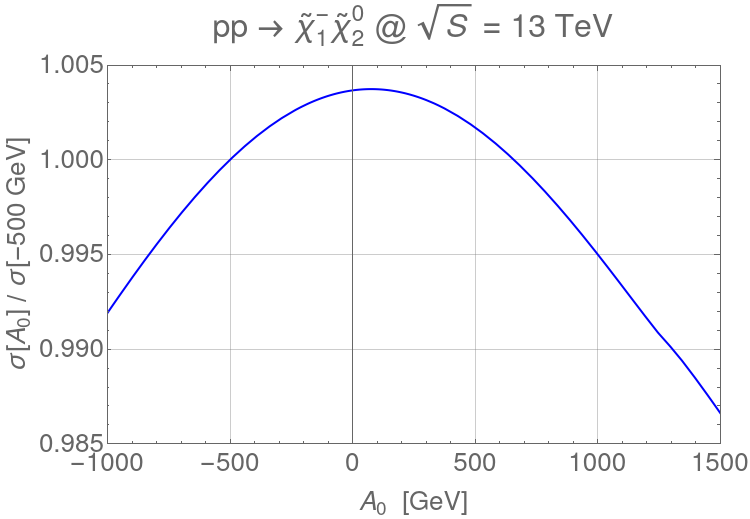}
\caption{Dependence of the NLO (or NLO+NLL or aNNLO+NNLL) total cross
  section on the common trilinear coupling $A_0$ that governs squark
  mixing in the sbottom sector.}
\label{fig:8}
\end{center}
\end{figure}

When the squark masses are not all identified with each other, but mixing in the
sbottom sector is allowed, a dependence on the trilinear coupling $A_0$ is
introduced. It is shown in Fig.\ \ref{fig:8}. As expected, for higgsinos it
is also weak and amounts to at most $+0.4$\% and $-1.3$\% when compared with
the cross section in our default scenario with $A_0 =-500$ GeV.

\section{Numerical results for gaugino pair production}
\label{sec:4}

We now turn to the case where the produced neutralinos and charginos have a large
gaugino component. The next-to-lightest neutralino $\tilde{\chi}^0_2$ and the
charginos $\tilde{\chi}^\pm_1$ will be considered as wino-like and almost
degenerate with a mass above 1100 GeV to satisfy experimental constraints, while
the LSP $\tilde{\chi}^0_1$ is assumed to be bino-like and light. In this scenario,
large production cross sections of $\tilde{\chi}^\pm_1\tilde{\chi}^0_2$ and
short decay chains are expected. For example, assuming an intermediate and equal
mass for left-handed staus and tau sneutrinos, the winos will decay through these
states into the LSP, taus and tau neutrinos, leading to interesting collider
signatures \cite{Aaboud:2017nhr}. This particular spectrum of particle masses can
be achieved within the phenomenological MSSM (pMSSM) framework. It is of
particular interest, since the coannihilation of light staus with the LSP can
generate a dark matter relic density in accordance with the observations
\cite{Ellis:1999mm,Branahl:2019yot}.

Based on an integrated LHC luminosity of 36 fb$^{-1}$, the ATLAS (CMS)
collaboration have excluded pure, mass-degenerate wino pairs $\ca\nb$ ($\cpa\cma$)
decaying with 100\% branching ratio via sleptons to significantly lighter pure
binos $\na$ 
up to masses of 1100 (800) GeV \cite{Aaboud:2018jiw,Sirunyan:2018lul}. With 139
fb$^{-1}$, the ATLAS collaboration could also exclude chargino pairs up to masses
of 1000 GeV \cite{Aad:2019vnb}. For pure winos decaying to on-shell gauge and
Higgs bosons, the ATLAS (CMS) limits are sometimes considerably weaker and reach
only 345 to 1000 (650) GeV, depending on the analysis method and despite
luminosities of up to 139 fb$^{-1}$ \cite{Aad:2019vvi,Aad:2019vvf,Aad:2019vnb,%
Sirunyan:2018ubx}.

The dependence on the gluino mass, which enters only at NLO, is again expected
to be weak. However, the squark mass dependence will now be more important, as
heavy gauginos can have large LO contributions from $t$- and $u$-channel diagrams
and their (negative) interferences with the $s$-channel. In addition, hadronic
gaugino decay channels will be open when $m_{\tilde{q}}<m_{\ca,\nb}$, and squark
threshold effects will appear in the one-loop diagrams when $m_{\ca,\nb}\simeq
m_{\tilde{q}}$. These thresholds will also affect the dependence on the trilinear
coupling $A_0$ controlling the physical sbottom masses.

Our desired SUSY spectrum with wino-like charginos and neutralinos and a
bino-like LSP is obtained using again the public code SPheno 4.0.3
\cite{Porod:2003um,Porod:2011nf} and by choosing a small value for the bino
mass parameter $M_1=100$ GeV, while the wino mass parameter $M_2>1$ TeV is chosen above
the ATLAS exclusion limits. The large gaugino content can be achieved by
choosing a large value for $\mu=3$ TeV $ \gg M_2$. With this
configuration, only a very small splitting between the masses of the
neutralino $\tilde{\chi}^0_2$ and the charginos $\tilde{\chi}^\pm_1$ is
generated, while the LSP $\na$ remains light.

\subsection{Invariant-mass distributions}

\begin{figure}
\begin{center}
\includegraphics[width=0.75\textwidth]{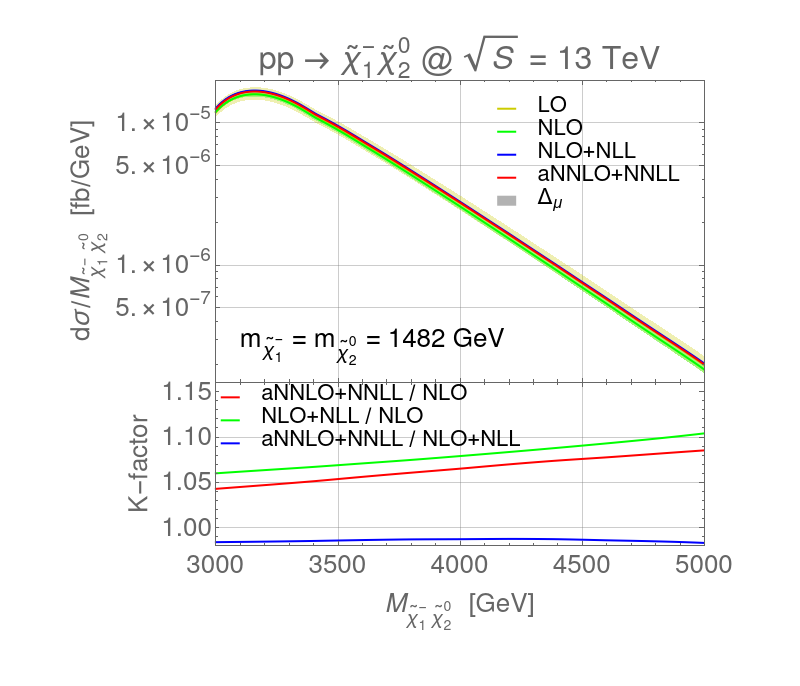}
\caption{Top: Invariant-mass distribution for the associated production of 
  charginos and neutralinos with masses of 1482 GeV
  at the LHC with a center-of-mass energy of $\sqrt{S}=13$~TeV. Shown are results
  at LO (yellow), NLO (green), NLO+NLL (blue) and aNNLO+NNLL (red) together with
  the corresponding scale uncertainties (shaded bands). Bottom: Ratios ($K$
  factors) of aNNLO+NNLL over NLO (red), NLO+NLL over NLO (green) and aNNLO+NNLL
  over NLO+NLL (blue) differential cross sections as a function of the invariant
  mass of the gaugino pair.}
\label{fig:9}
\end{center}
\end{figure}

We begin our discussion with the invariant-mass distribution for the associated
production of wino-like lightest charginos and second-lightest neutralinos. These
differential cross sections at LO (yellow), NLO (green), NLO+NLL (blue) and aNNLO+NLL
(red curve) are shown in the upper panel of Fig.~\ref{fig:9}. In contrast to the
higgsino case, the NLO corrections increase the LO cross section only at low
invariant masses, but decrease it for large invariant masses. A decrease for all
invariant masses is observed from NLO+NLL to aNNLO+NNLL (see also the lower panel).
This behavior is correlated with large $t$- and $u$-channel contributions and large
cancellations of the squared $s$-channel contribution with its interference terms. 

\begin{figure}
\begin{center}
\includegraphics[width=0.75\textwidth]{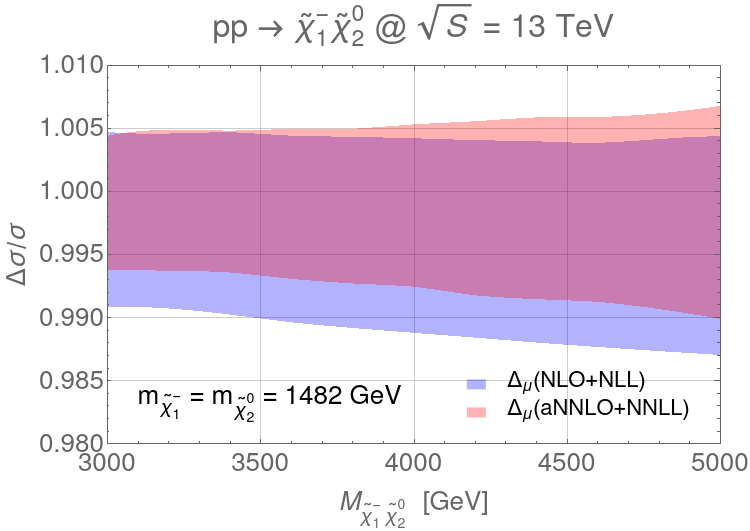}
\caption{Scale uncertainty of the invariant-mass distribution in Fig.\ \ref{fig:9}.
  Shown are the results at NLO+NLL (blue) and aNNLO+NNLL (red shaded band).}
\label{fig:10}
\end{center}
\end{figure}

The combined scale uncertainty for this distribution is shown in Fig.\ \ref{fig:10}
at NLO+NLL (blue) and aNNLO+NNLL (red). A reduction from $\pm0.7$\% to $\pm 0.5$\%
is observed at low invariant masses. The reduction is smaller for large invariant
masses, which is again related to the importance of the $t$- and $u$-channels.

\subsection{Total cross sections}

\begin{figure}
\begin{center}
\includegraphics[width=0.75\textwidth]{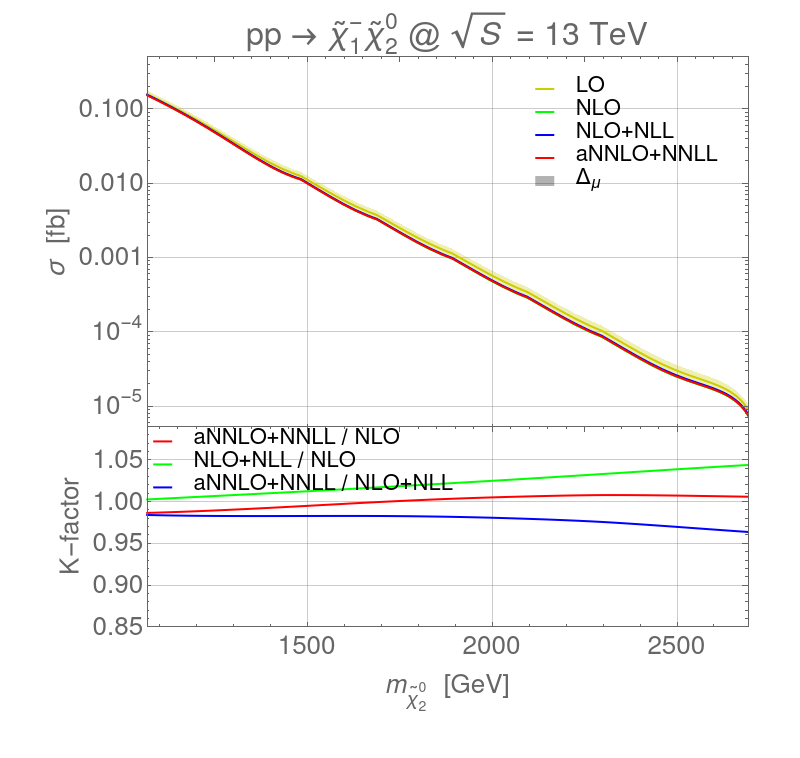}
\caption{Top: Total cross section for gaugino-like charginos and neutralinos at
  the LHC with a center-of-mass energy
  of $\sqrt{S}=13$ TeV as a function of the $\nb$ mass. Shown are results at LO
  (yellow), NLO (green), NLO+NLL (blue) and aNNLO+NNLL (red) together with the
  corresponding scale uncertainties (shaded bands). Bottom: Ratios ($K$
  factors) of aNNLO+NNLL over NLO (red), NLO+NLL over NLO (green) and aNNLO+NNLL
  over NLO+NLL (blue) total cross sections.}
\label{fig:11}
\end{center}
\end{figure}

We now turn to the total cross sections for gauginos. They are shown in Fig.\ \ref{fig:11}
for the associated production of a negatively charged and a neutral wino
as a function of the second-lightest neutralino (and lightest chargino) mass.
For our choice of parameters and after integration over the invariant mass, the
NLO prediction is smaller than the LO one over the entire $\nb$ mass range,
and the size of the aNNLO+NNLL corrections is very small in this particular case.

\begin{figure}
\begin{center}
\includegraphics[width=0.49\textwidth]{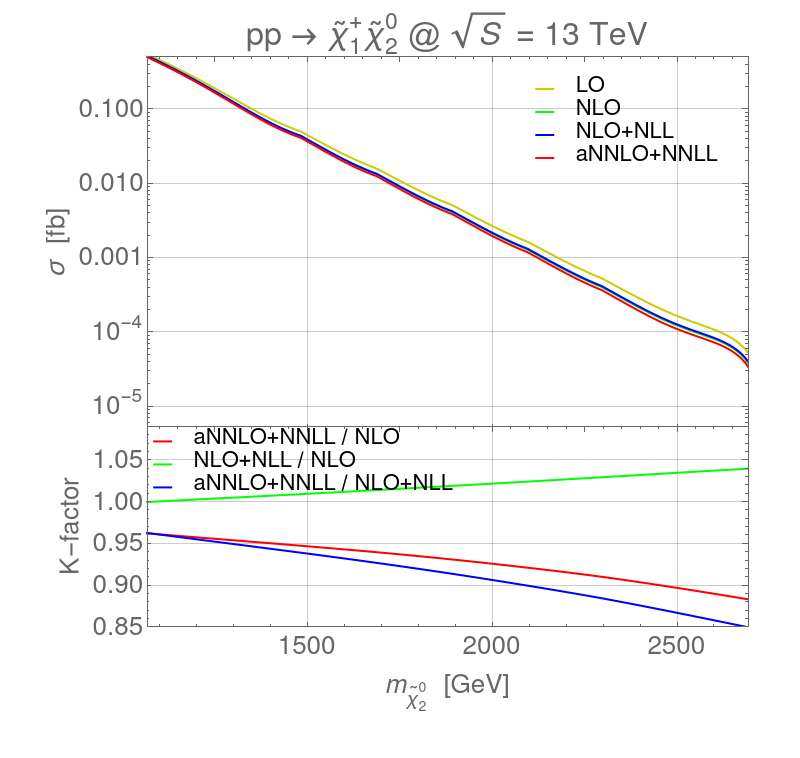}
\includegraphics[width=0.49\textwidth]{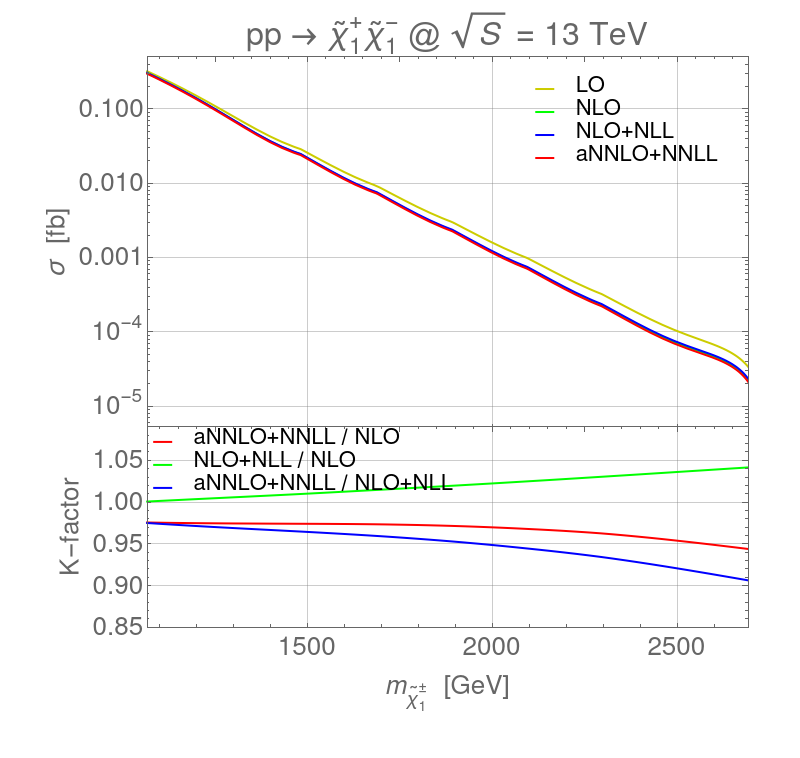}
\caption{Same as Fig.\ \ref{fig:11}, but for the associated production of a positively
  charged gaugino with the second-lightest neutralinos (left) and for the pair
  production of charginos (right).}
\label{fig:12}
\end{center}
\end{figure}

This is, however, a peculiarity of the chosen channel with a negative chargino,
as can be seen from Fig.\ \ref{fig:12} showing the total cross sections for the
associated production of a positive chargino with a neutralino (left) and for
chargino pair production (right).
Both the absolute size of the cross section and the size of the corrections are
then different due to the fact that we probe large momentum fractions $x$ and
therefore the valence quark structure in the PDFs. In particular, the cross
section for $\cpa\nb$ is larger than the one for $\cma\nb$ by about a factor of
four, and the aNNLO+NNLL corrections now amount to up to -12 to -15\% with respect
to the NLO and NLO+NLL predictions. The cross section for chargino pair production
through a neutral current represents an intermediate case, as expected.

\begin{figure}
\begin{center}
\includegraphics[width=0.49\textwidth]{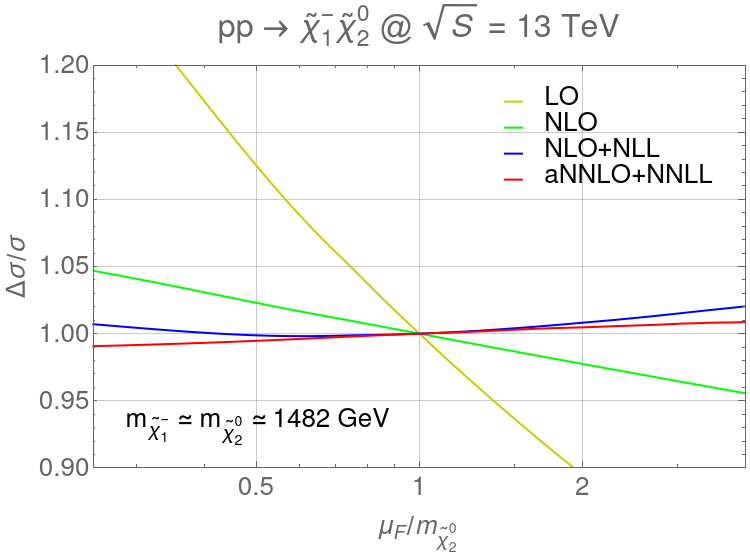}
\includegraphics[width=0.49\textwidth]{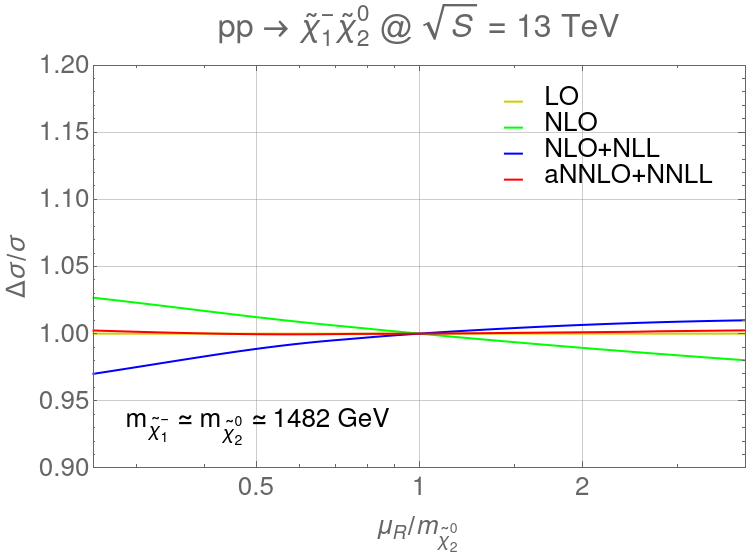}
\caption{Relative variation of the total cross section for gaugino-like
  chargino-neutralino pairs as a function of the factorization (top) and renormalization
  scale (bottom). Shown are results at LO (yellow), NLO (green), NLO+NLL
  (blue) and aNNLO+NNLL (red).}
\label{fig:13}
\end{center}
\end{figure}

The dependence of the total gaugino cross section on the factorization (top)
and renormalization (bottom) scale is studied individually in Fig.\ \ref{fig:13}.
As in the higggsino case, the LO cross section is independent of the
renormalization scale. The dependence introduced at NLO of 5\%
is reduced at NLO+NLL to 4\% and to below percent level at aNNLO+NNLL.
A similarly impressive reduction has been observed for sleptons
\cite{Fiaschi:2019zgh}.
The LO factorization scale dependence is much stronger than in the
higgsino case, as we are probing the evolution of the PDFs from the GeV-
to the TeV-region. It is reduced from more than 30\% at LO to 10\% at NLO,
then to 2 \% at NLO+NLL and aNNLO+NNLL. 

\begin{figure}
\begin{center}
\includegraphics[width=0.75\textwidth]{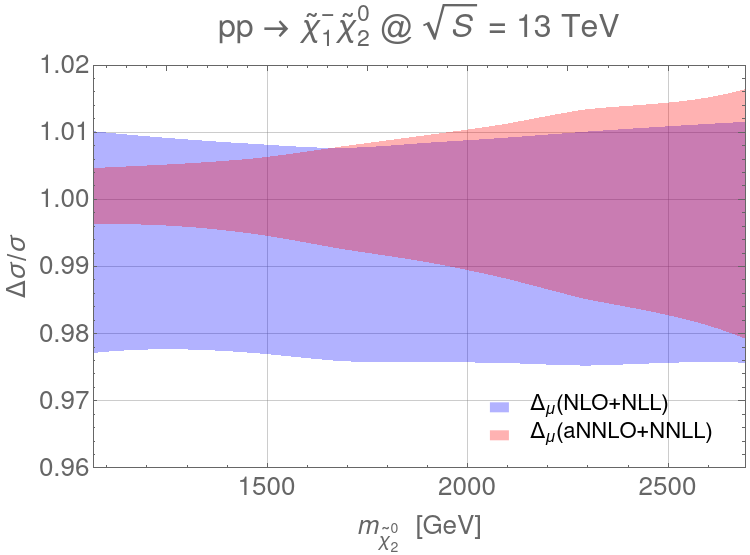}
\caption{Scale uncertainty of the total cross section for gaugino-like
  chargino-neutralino pairs at the LHC with a center-of-mass energy of
  $\sqrt{S}=13$ TeV as a function of the neutralino mass. Shown are the
  results at NLO+NLL (blue) and aNNLO+NNLL (red shaded band).}
\label{fig:14}
\end{center}
\end{figure}

In Fig.\ \ref{fig:11} we observed a better convergence of the perturbative
series for not too heavy gauginos than for very large masses. This behavior
is reflected in Fig.\ \ref{fig:14}, where the total scale uncertainty also
increases towards very large gaugino masses. At 1.1 TeV, it amounts to 3\%
at NLO+NLL and only 1\% at aNNLO+NNLL, while at 2.7 TeV it amounts to
3\% in both cases.\\

\begin{figure}
\begin{center}
\includegraphics[width=0.75\textwidth]{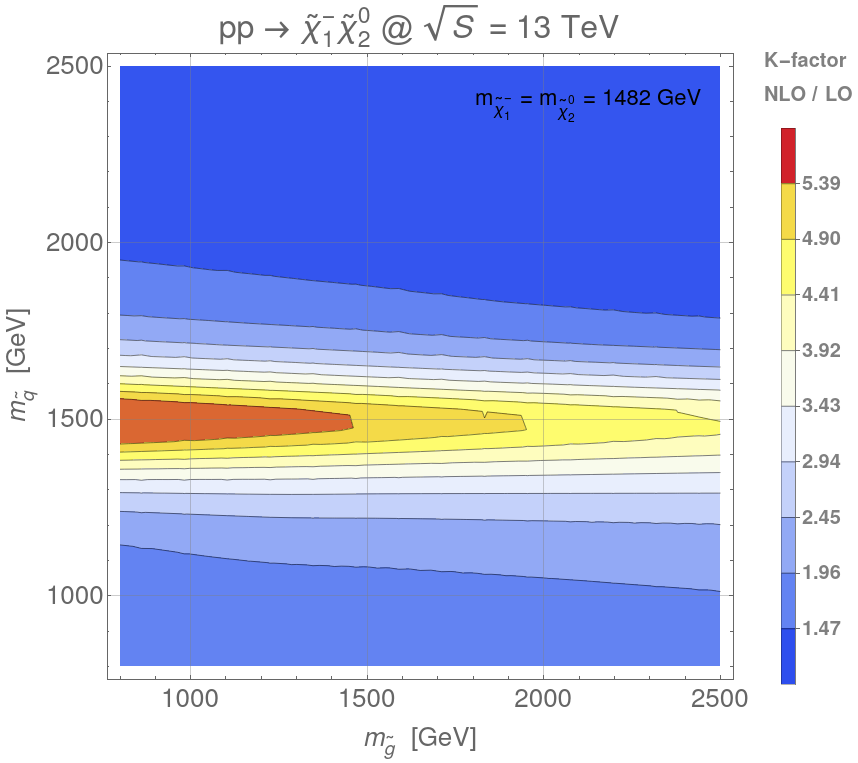}
\caption{Ratio ($K$ factor) of NLO over LO total cross sections (both
  with NLO PDFs) for gaugino pair production
  at the LHC with a center-of-mass energy of $\sqrt{S}=13$ TeV as a
  function of the squark and gluino masses.}
\label{fig:15}
\end{center}
\end{figure}

Fig.\ \ref{fig:15} shows the dependence of the NLO/LO $K$-factor for
the production of gaugino-like charginos and neutralinos on the
squark and gluino masses. As expected, the dependence on the gluino mass,
which enters only at NLO, is indeed weak and almost invisible, when the
squark mass differs substantially from the gaugino mass of about 1.5 TeV.
In contrast, when the squark mass is close to the gaugino mass, the squark mass
has a substantial influence already at tree-level, but also at NLO (and
beyond), when the squark threshold is crossed in virtual box diagrams. In
this situation, also the gluino mass can induce a significant variation
of the $K$-factor. The fact that the NLO/LO cross section ratio can reach
values much larger than one is related to the (almost) on-shell production
of intermediate squarks in the final state that subsequently decay into
the observed gauginos. This
situation therefore requires a careful identification of squark and gaugino
production, respectively, from the observed decay products and in particular
the presence of jets.

\begin{figure}
\begin{center}
\includegraphics[width=0.75\textwidth]{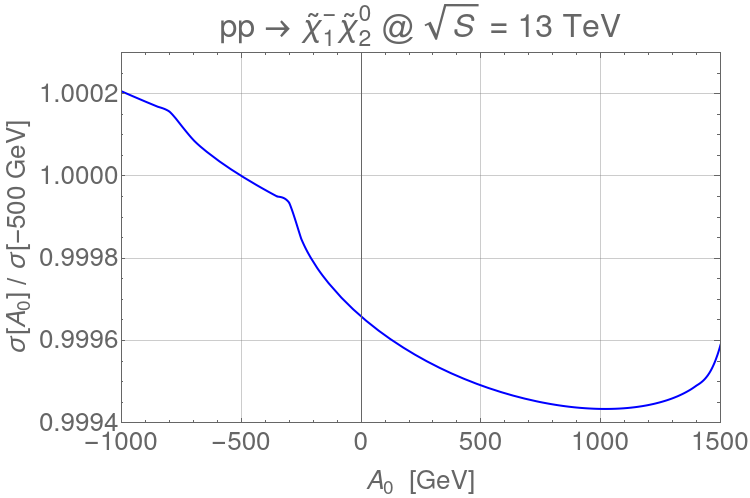}
\caption{Dependence of the NLO (or NLO+NLL or aNNLO+NNLL) total cross
  section on the common trilinear coupling $A_0$ that governs squark
  mixing in the sbottom sector. Shown is the ratio over the default
  scenario with $A_0=-500$ GeV.}
\label{fig:16}
\end{center}
\end{figure}

The presence of squark thresholds is also observed in Fig.\ \ref{fig:16}
for bottom squarks. This figure shows the dependence of the NLO (or NLO+NLL
or aNNLO+NNLL) total gaugino cross section on the trilinear coupling $A_0$
over its value for our default choice of $A_0=-500$ GeV. While the overall
dependence is very weak, as bottom quarks in the proton PDFs contribute very
little to the total cross section at these large values of $x$, the kinks
when the two physical sbottom mass thresholds are crossed are nevertheless
clearly visible at $A_0=-800$ GeV and $A_0=-300$ GeV.

\section{Conclusion}
\label{sec:5}

In conclusion, we have presented in this paper the improvement of our
previous predictions for higgsino and gaugino pair production at the LHC
from NLO+NLL to aNNLO+NNLL precision. We have briefly reviewed the formalism
employed for the threshold resummation of large logarithms that can spoil
the convergence of the perturbative series and highlighted the analytical
results required for the resummation at NNLL accuracy and its matching to
the fixed order calculation at aNNLO. Numerical results were presented for
two very different scenarios, i.e. higgsino and gaugino pair production at
the LHC. The mass limits on higgsinos from the LHC are still relatively weak,
they can thus still be as light as a (few) hundred GeV and consequently
produced mostly in the $s$-channel. The aNNLO+NNLL results were found to
induce only small modifications of the differential and total cross sections
and to stabilize them even more than before at NLO+NLL with respect to
variations of the factorization and renormalization scales. For gauginos,
which like squarks and gluinos have recently been constrained by LHC searches
to the TeV region and beyond, also $t$- and $u$-channels and thus the
dependence on the squark mass became important already at tree-level, and
the impact of the higher-order corrections in the large $x$-region required
a closer look. It varied not only with the considered production channel,
i.e.\ the total charge of the final state, but also with the squark mass and,
in the threshold region, even the gluino mass. As an additional new aspect,
we included in our calculation explicitly the mixing in the squark sector,
which proved to be relevant in practice only for bottom (s)quarks and thus
more for light higgsinos produced from partons at small $x$ than for
heavier gauginos produced from partons at larger values of $x$.

\begin{acknowledgements}
This work has been supported by the BMBF under contract 05H18PMCC1.
\end{acknowledgements}

\bibliography{bib}

\end{document}